\documentclass[onecolumn,superscriptaddress,aps]{revtex4}
\usepackage[T1]{fontenc}
\usepackage[english]{babel}
\usepackage{amsmath}
\usepackage{amssymb}
\usepackage{graphicx}
\usepackage{bm}
\usepackage{units}
\usepackage{braket}
\usepackage{float}
\usepackage{textcomp}
\usepackage{color}
\usepackage{multirow}
\usepackage[normalem]{ulem}

\graphicspath{{images/}}

\begin{document}

\author{Gabriel Dufour}
\email{gabriel.dufour@physik.uni-freiburg.de}
\affiliation{Physikalisches Institut, Albert-Ludwigs-Universit\''at Freiburg, Hermann-Herder-Stra\ss e 3, 79104 Freiburg, Germany}

\author{Tobias Br\"unner}
\email{tobias.bruenner@physik.uni-freiburg.de}
\affiliation{Physikalisches Institut, Albert-Ludwigs-Universit\''at Freiburg, Hermann-Herder-Stra\ss e 3, 79104 Freiburg, Germany}

\author{Christoph Dittel}
\affiliation{Institut f\"ur Experimentalphysik, Universit\"at Innsbruck, Technikerstra\ss e 25, 6020 Innsbruck, Austria}

\author{Gregor Weihs}
\affiliation{Institut f\"ur Experimentalphysik, Universit\"at Innsbruck, Technikerstra\ss e 25, 6020 Innsbruck, Austria}

\author{Robert Keil}
\affiliation{Institut f\"ur Experimentalphysik, Universit\"at Innsbruck, Technikerstra\ss e 25, 6020 Innsbruck, Austria}

\author{Andreas Buchleitner}
\affiliation{Physikalisches Institut, Albert-Ludwigs-Universität Freiburg, Hermann-Herder-Stra\ss e 3, 79104 Freiburg, Germany}

\title{Many-particle interference in a two-component bosonic Josephson junction:\\ an all-optical simulation}

\begin{abstract}
We conceive an all-optical representation of the dynamics of two distinct types of interacting bosons in a double well by an array of evanescently coupled photonic waveguides. Many-particle interference effects are probed for various interaction strengths by changing the relative abundance of the particle species and can be readily identified by monitoring the propagation of the light intensity across the waveguide array. In particular, we show that finite inter-particle interaction strengths reduce the many-particle interference contrast by dephasing. A general description of the many-particle dynamics for arbitrary initial states is given in terms of two coupled spins by generalising the Schwinger representation to two particle species.
\end{abstract}

\date{\today}
\maketitle

\section{Introduction}

The dynamics of a many-body quantum system crucially depends on its constituents' mutual distinguishability -- which determines which paths relating initial and final states are allowed to interfere.
This is illustrated by the paradigmatic Hong-Ou-Mandel (HOM) experiment, where two photons impinge on opposite input ports of a 
balanced beam splitter \cite{hong_measurement_1987}.  
 If the photons can be distinguished, for example by their arrival time,
the probability of observing one photon in each output port will be equal to the probability that both photons are transmitted plus the 
probability that both are reflected, that is 50\%.
On the other hand, if the photons are indistinguishable, the two alternative two-particle paths interfere destructively, and the associated coincidence events are completely suppressed. 
A continuous transition between both extreme cases can be induced by continuous tuning of the photons' relative arrival times.
To date, the impact of (in)distinguishability on the interference of many non-interacting particles has 
been generalized to highly symmetric \cite{campos89,belinskii92,lim2005,tichy2010,tichy2011,mayer2011,tichy2012,tichy2014,crespi_suppression_2015,dittel_many-body_2017} and random 
\cite{beenakker2009,schlawin2012,aaronson_computational_2013,walschaers_annrev16,mat_diss} multimode scattering scenarios, where 
bosonic many-particle
input states are transmitted across a multimode scattering device. A versatile means to realize these scenarios is offered by networks of passive optical elements \cite{Reck-ER-1994}, i.e. beam splitters and phase shifters, generalizing the HOM setup to an increasing 
number of non-interacting particles and modes. As in the HOM case, the output of such a device is predicted to depend on the distinguishability 
of the involved particles, 
as well as on specific features of the scatterer \cite{tichy_interference_2014,Shchesnovich-SC-2014,shchesnovich_partial_2015,tichy_sampling_2015,Tillmann:PartialDistinguishability,walschaers_statistical_2016,tamma_multi-boson_2016,urbina_multiparticle_2016}, and rich experimental evidence thereof has been reported \cite{Ra_PNAS13,Ra:DetectiondependentCoherenceTimes,Broome:BosonSampling,Spring:BosonSampling,Tillmann:BosonSampling,Crespi:BosonSampling,metcalf_multiphoton_2013,Carolan:VerfificationBosonSampling,Spagnolo:VerificationBosonSampling,menssen17}.

But how general are such indistinguishability-induced many-particle interference phenomena, given that, in generic many-particle systems, particles are also coupled by  inter-particle interactions -- a
much more ``classical'' mechanism as compared to indistinguishability? For example, ensembles of ultracold atoms are reputed to display ``new physics'' -- but which of their
dynamical features are due to the particles' mutual indistinguishability, and which are caused by their
interactions? 
As a first step to discriminate these 
distinct sources of non-trivial many-particle quantum dynamics, we here take a closer look at an interacting many-particle 
generalization of the HOM setup, where the beam splitter is replaced by a bosonic Josephson junction (BJJ) which couples two modes populated by interacting bosons \cite{smerzi_quantum_1997,milburn_quantum_1997,dalton_two_2012}. 
Originally inspired by the scenario of two superconductors coupled by a thin insulating layer \cite{josephson_possible_1962}, Josephson junctions have since been realized by coupling two superfluid helium reservoirs through nanometric apertures \cite{pereverzev_quantum_1997,sukhatme_observation_2001} and with Bose-Einstein condensates trapped in double-well potentials \cite{albiez_direct_2005,gati_bosonic_2007}.
We introduce distinguishability in the BJJ by considering mixtures of two mutually distinguishable particle species, and propose an all-optical simulation of the thus defined two-component BJJ by a two-dimensional (2D) array of photonic waveguides. 
Such waveguide lattices are a long-established tool to simulate elusive single-particle phenomena in a controlled environment \cite{Longhi:Quantum_Analogies_Review,Keil:GlauberFockLattices,Verbin:QuasicrystalPhaseTransition,Weimann:SurfaceFanoState,Plotnik:UnconventionalEdgeState} but they can also be used to study systems with light-matter interaction \cite{Crespi:QuantumRabiModel} or pairs of identical interacting particles \cite{lahini_quantum_2012,Corrielli2013,rai_photonic_2015,Mukherjee:CDTInteractingPair}. Inspired by a proposal to optically simulate a single-component BJJ in a planar array \cite{longhi_optical_2011}, in our setting with two particle types, single-particle tunnelling between the 
potential wells is implemented by engineered evanescent couplings between the waveguides, inter-particle interactions are induced by refractive 
index variations from waveguide to waveguide, and the two distinct particle types are represented by the two dimensions of the lattice. For our formal 
treatment of the problem, we employ the Schwinger representation \cite{schwinger_angular_1952} to map the BJJ Hamiltonian onto that of two interacting spins, each representing the state of either particle species, thereby exploiting the mathematical tools associated with the SU(2) group. We then show that even if both particle species experience identical potentials, the system's behaviour strongly depends on the way the particles are allotted among the species, because this partition determines those interference processes which manifest in the subsequent time evolution. We also show how a 
non-vanishing interaction strength between the particles tends to wash out these interferences. Our results add a new perspective on the 
widely explored dynamics of the two-component BJJ, which hitherto focused on
mean field dynamics \cite{xu_stability_2008,julia-diaz_josephson_2009,mazzarella_atomic_2009,tichy_dynamical_2012}, on the competition between intra- and inter-species interactions \cite{zin_quantum_2011,citro_quantum_2011}, on the generation and characterization of entanglement \cite{ng_quantum-correlated_2003,mujal_quantum_2016},
and on synchronization effects \cite{qiu_hybrid_2015}.

The paper is structured as follows: In section \ref{Sec:Model}, we define the two-component BJJ and establish its mapping onto a lattice of coupled optical waveguides. In section \ref{Sec:Schwinger}, we reformulate the problem in terms of two interacting spins 
and introduce basis states associated with the total spin.
Section \ref{Sec:Typical} contains a discussion of realistic parameters for an experimental realization of the waveguide lattice. In section \ref{Sec:Cases}, we compare the time evolution of the intensity distribution across the waveguide lattice, for initial states corresponding to the same total particle number in each well of the BJJ model, but with different repartitions of both particle types.
Section \ref{Conc} concludes the paper.

\section{Model and its optical realisation}
\label{Sec:Model}
\begin{figure}[t]
	\includegraphics[width=\columnwidth]{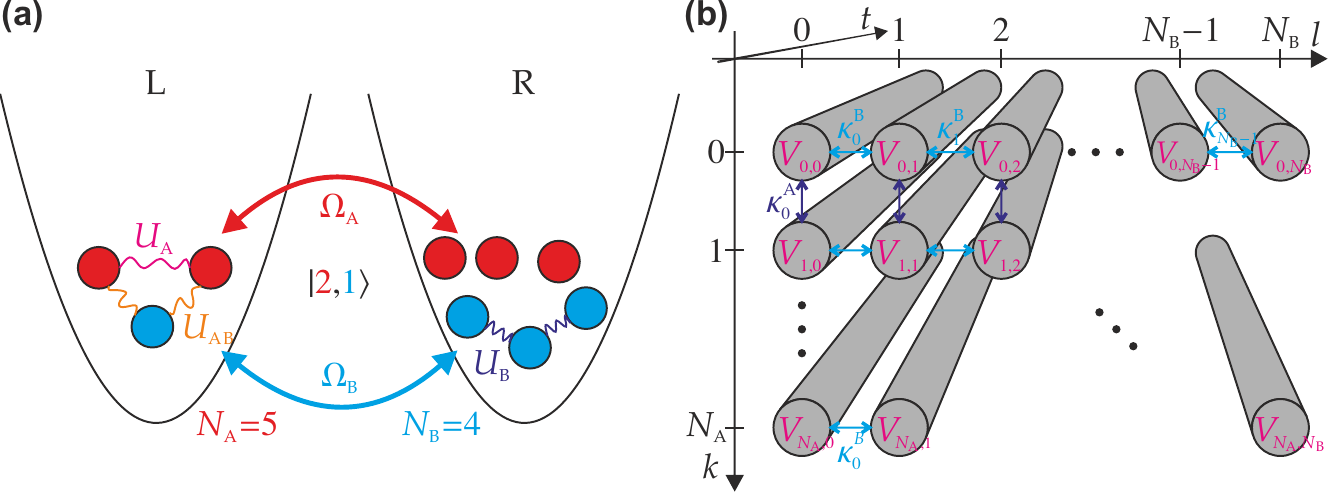}
	\caption{\label{Fig:DoubleWell} Two-component BJJ and its 	all-optical analogue. \textbf{(a)} Two bosonic species in a double-well potential. Particles of type A (B) tunnel between the two sites L and R with a rate $\Omega_\mathrm{A}$ ($\Omega_\mathrm{B}$), and  interact with particles of the same species with strength $U_\mathrm{A}$ ($U_\mathrm{B}$). The interspecies interaction strength is denoted by $U_\mathrm{AB}$. \textbf{(b)} Tight-binding  lattice of $(N_\mathrm{A}+1)\times (N_\mathrm{B}+1)$ waveguides. Light is propagating along the $t$-axis. The vertical (horizontal) coupling strength between neighbouring sites is given by $\kappa^\mathrm{A}_k$ ($\kappa^\mathrm{B}_l$), while the individual site energies are detuned by $V_{k,l}$. In an idealized implementation, coupling occurs only in the vertical and horizontal directions. For a discussion of undesired couplings along diagonal directions, see Secs.~IV and V of the main text.}
\end{figure}

Let us consider $N_\mathrm{A}$ bosons of type $A$ and $N_\mathrm{B}=N-N_\mathrm{A}$ bosons of type $B$ in a symmetric double-well potential, or bosonic Josephson junction (see Fig.~\ref{Fig:DoubleWell} \textbf{(a)}). 
In the tight-binding approximation, we consider only two modes, L and R, localized in the left and right well, respectively. Particles of type A (B) tunnel between these two modes with rates $\Omega_\mathrm{A}$ ($\Omega_\mathrm{B}$). 	
Thus, a single type-B particle initialized in the left well at time $t=0$ has tunnelled to the right well at time $t=\pi/\Omega_\mathrm{B}$ and is back in the left well at time $t=2\pi/\Omega_\mathrm{B}$. At times $t=\pi/(2\Omega_\mathrm{B})$ and $t=3\pi/(2\Omega_\mathrm{B})$, it is in a balanced superposition of both modes. 
The two wells at the initial (final) time can hence be identified with the two input (output) modes of a beam splitter with a reflectivity depending on the evolution time: perfectly reflecting at times $t=0,\  2\pi/\Omega_\mathrm{B},\ \dots$, perfectly transparent at times  $t=\pi/\Omega_\mathrm{B},\ 3\pi/\Omega_\mathrm{B} ,\ \dots$ and balanced at times $t=\pi/(2\Omega_\mathrm{B}),\ 3\pi/(2\Omega_\mathrm{B}),\ \dots$ .
For several particles, however, unlike what can be realized with photons interfering on a beam splitter,
we also include on-site interactions of strength	$U_\mathrm{A}$ ($ U_\mathrm{B}$) between type A (B) particles and of strength $U_\mathrm{AB}$  between A and B particles.
 The system is thus described by the Bose-Hubbard Hamiltonian with two sites \cite{mujal_quantum_2016}, 
\begin{align}
\hat{H}= & -\frac{\hbar \Omega_\mathrm{A}}{2} (\hat{a}_\mathrm{L}^\dagger \hat{a}_\mathrm{R}+\hat{a}_\mathrm{R}^\dagger \hat{a}_\mathrm{L}) +\frac{\hbar U_\mathrm{A}}{2}
 \left(   \hat{a}_\mathrm{L}^{\dagger 2} 
  \hat{a}_\mathrm{L}^2+\hat{a}_\mathrm{R}^{\dagger 2} \hat{a}_\mathrm{R}^2 \right)\notag \\
& -\frac{\hbar \Omega_\mathrm{B}}{2} (\hat{b}_\mathrm{L}^\dagger \hat{b}_\mathrm{R}+\hat{b}_\mathrm{R}^\dagger \hat{b}_\mathrm{L}) +\frac{\hbar U_\mathrm{B}}{2} \left(   \hat{b}_\mathrm{L}^{\dagger 2} \hat{b}_\mathrm{L}^2+\hat{b}_\mathrm{R}^{\dagger 2} \hat{b}_\mathrm{R}^2 \right)\notag \\
 &+\hbar U_\mathrm{AB} \left(\hat{a}_\mathrm{L}^{\dagger} \hat{a}_\mathrm{L} \hat{b}_\mathrm{L}^{\dagger} \hat{b}_\mathrm{L}+\hat{a}_\mathrm{R}^{\dagger} \hat{a}_\mathrm{R} \hat{b}_\mathrm{R}^{\dagger} \hat{b}_\mathrm{R}\right),
\label{Hamiltonian}
\end{align}
with $\hat a_\ell\ (\hat b_\ell)$ the annihilation operator of particles of species A (B) at site $\ell\in\{\mathrm{L},\mathrm{R}\}$.
Particles of the same kind are mutually indistinguishable, while particles of different types are distinguishable from one another. Hence, creation and annihilation operators associated with the same particle species obey the usual bosonic commutation relations, while those belonging to different species always commute. 
In the special case where $\Omega_\mathrm{A}=\Omega_\mathrm{B}=\Omega$ and $U_\mathrm{A}=U_\mathrm{B}=U_\mathrm{AB}=U$, so that all particles have the same tunnelling rate $\Omega$ and interaction strength $U$, irrespective of their type, the Hamiltonian is termed \emph{isospecific} \cite{tichy_dynamical_2012}.

To map the  quantum dynamics of this system of $N$ interacting particles to the propagation of coherent light in an array of waveguides, we generalize Longhi's approach  \cite{longhi_optical_2011} by defining the basis functions
\begin{equation}
\ket{\Psi(t)}\equiv\sum_{k=0}^{N_\mathrm{A}}\sum_{l=0}^{N_\mathrm{B}}c_{k,l}(t)\ket{k,l},
\end{equation}
where $\ket{k,l}$ denotes the Fock state with $k$ particles of type A and $l$ particles of type B in the left mode, and  all remaining particles in the right mode (see Fig.~\ref{Fig:DoubleWell} \textbf{(a)}), such that
\begin{equation}
\ket{k,l}\propto \left( \hat{a}_\mathrm{L}^{\dagger}\right) ^k \left( \hat{a}_\mathrm{R}^{\dagger}\right) ^{N_\mathrm{A}-k}\left( \hat{b}_\mathrm{L}^{\dagger}\right) ^l\left( \hat{b}_\mathrm{R}^{\dagger}\right) ^{N_\mathrm{B}-l}\ket{0}.
\end{equation}
Inserting this ansatz for $\ket{\Psi(t)}$ into the Schr\"odinger equation generated by $\hat{H}$ yields the coupled differential equations
\begin{equation}
i\frac{\mathrm{d}c_{k,l}}{\mathrm{d}t}+\kappa_{k}^\mathrm{A} c_{k+1,l}(t)+\kappa_{k-1}^\mathrm{A} c_{k-1,l}(t)+\kappa_{l}^\mathrm{B} c_{k,l+1}(t)+\kappa_{l-1}^\mathrm{B} c_{k,l-1}(t)+V_{k,l}c_{k,l}(t)=0,
\label{TightBinding}
\end{equation}
for all $k=0,\ldots,N_\mathrm{A}$ and $l=0,\ldots,N_\mathrm{B}$, where 
\begin{equation}
V_{k,l}= -\frac{U_\mathrm{A}}{2}\left( k^2+ (N_\mathrm{A}-k)^2 -N_\mathrm{A}      \right)      -\frac{U_\mathrm{B}}{2}\left( l^2+ (N_\mathrm{B}-l)^2 -N_\mathrm{B}      \right)  
-U_{AB} \left( kl+ (N_\mathrm{A}-k)(N_\mathrm{B}-l)      \right)  
\label{detuning}
\end{equation}
and
\begin{align}
\kappa_k^\mathrm{A}&=\frac{\Omega_\mathrm{A}}{2} \sqrt{(k+1)(N_\mathrm{A}-k)}, & \kappa_l^\mathrm{B}&=\frac{\Omega_\mathrm{B}}{2}\sqrt{(l+1)(N_\mathrm{B}-l)}.
\label{coupling}
\end{align}
 Equation~\eqref{TightBinding} can be read as a tight-binding evolution equation for amplitudes defined on a square lattice of $(N_\mathrm{A}+1)\times (N_\mathrm{B}+1)$ sites 
 with nearest-neighbour couplings $\kappa_k^\mathrm{A}$ ($\kappa_l^\mathrm{B}$) in the vertical (horizontal)  direction, and on-site detuning $V_{k,l}$,  as sketched in Fig.~\ref{Fig:DoubleWell} \textbf{(b)}. 
 This physical scenario can be faithfully implemented by propagating light through a 2D array of parallel, evanescently coupled, single-mode waveguides.
  As illustrated in Fig.~\ref{Fig:DoubleWell} \textbf{(b)}, time $t$ is identified with the spatial coordinate along the waveguides, and in this picture, 
  $c_{k,l}(t)$ corresponds to the field amplitude at point $t$ in the waveguide with lattice coordinates $(k,l)$  \cite{szameit_light_2007,lederer_discrete_2008}.
 An initial state $\ket{\psi_0}$ is prepared by injecting field amplitudes $c_{k,l}(0)=\braket{k,l|\psi_0}$ in the corresponding waveguide modes. The probability of measuring the system in state $\ket{k,l}$ after an evolution time $t$ is 
 \begin{align}
p_{k,l}(t)=|\braket{k,l|\exp(-i\hat Ht/\hbar)|\psi_0}|^2=|c_{k,l}(t)|^2\, ,
\end{align}
 and is therefore given by the (normalized) intensity at a distance $t$ along the waveguide with coordinates $(k,l)$.
We define the total population imbalance between the two wells, irrespective of the species, as $m= k+l-N/2$,  such that there are $N/2+m$ particles in total in the left well, and $N/2-m$ in the right well.  For an even (odd) total particle number $N$, $m$ takes integer (half-integer) values between $-N/2$ and $N/2$. The probability to measure an imbalance $m$ reads
  \begin{align}\label{totimbalance}
p_m(t)=\sum_{k+l=m+N/2} p_{k,l}(t)=\sum_{k+l=m+N/2} |c_{k,l}(t)|^2,
\end{align}
which is the sum of light intensities along an antidiagonal of the array. 
While the probability distribution $p_{k,l}(t)$ directly resolves the distribution of the individual species, the total population imbalance  does not.  Notwithstanding, due to many-particle interference, we show that the evolution of $p_m(t)$ is strongly affected by the bi-component structure of the system, making it an appropriate observable to study the effect of (in)distinguishability on the dynamics.

\section{Schwinger spin representation}
\label{Sec:Schwinger}
 
It is instructive to reformulate the problem in terms of a pair of coupled spins, using the mapping between bosonic modes and spin operators introduced by Schwinger \cite{schwinger_angular_1952}.
The Hilbert space of a single particle is identical to that of a spin 1/2, where we  identify the state of the particle in the left (right) mode with spin up (down).  For two indistinguishable bosons, the symmetry of the wavefunction imposes that the two spins are in the triplet state. More generally, for $N$ indistinguishable bosons, the exchange symmetry ensures that the collective spin moves on a sphere of radius $N/2$. 
In the present case of two species, we can thus define two spin operators  $\hat{\vec{J}}_\mathrm{A}$ and $\hat{\vec{J}}_\mathrm{B}$ 
with magnitudes 
related to the particle numbers through
 	\begin{align}
 	\braket{\hat{J}_\mathrm{A}^2}&=\frac{N_\mathrm{A}}{2}\left(\frac{N_\mathrm{A}}{2}+1 \right), &\braket{\hat{J}_\mathrm{B}^2}&=\frac{N_\mathrm{B}}{2}\left(\frac{N_\mathrm{B}}{2}+1 \right).
 	\end{align}
The components of  $\hat{\vec{J}}_\mathrm{A}$ are explicitly given by
 \begin{subequations}
 \begin{align}
\hat{J}_{\mathrm{A}x}&=\frac12 \left(\hat{a}_\mathrm{L}^\dagger \hat{a}_\mathrm{R}+ \hat{a}_\mathrm{R}^\dagger \hat{a}_\mathrm{L}\right),\\ 
 \hat{J}_{\mathrm{A}y}&=\frac{1}{2i} \left(\hat{a}_\mathrm{L}^\dagger \hat{a}_\mathrm{R}- \hat{a}_\mathrm{R}^\dagger \hat{a}_\mathrm{L}\right),\\
 \hat{J}_{\mathrm{A}z}&=\frac12 \left(\hat{a}_\mathrm{L}^\dagger \hat{a}_\mathrm{L}- \hat{a}_\mathrm{R}^\dagger \hat{a}_\mathrm{R}\right)\, ,
 \end{align}
 \end{subequations}
and those of $\hat{\vec{J}}_\mathrm{B}$ analogously, in terms of creation and annihilation operators for type-B particles.
In particular, the spins' $z$-components measure the population imbalance between the two wells, for each species.
 The Fock state $\ket{k,l}$ can thus be rewritten as the common eigenstate $\ket{j_\mathrm{A},m_\mathrm{A},j_\mathrm{B},m_\mathrm{B}}$ of $\hat{J}_\mathrm{A}^2,\ \hat{J}_{\mathrm{A}z},\  \hat{J}_\mathrm{B}^2$ and $\hat{J}_{\mathrm{B}z}$, where we identify
  $j_\mathrm{A}=N_\mathrm{A}/2$, $m_\mathrm{A}=k-N_\mathrm{A}/2$, $j_\mathrm{B}=N_\mathrm{B}/2$ and $m_\mathrm{B}=l-N_\mathrm{B}/2$.
  Except for constant terms, the Hamiltonian \eqref{Hamiltonian} can therefore be restated as
  \begin{align}\label{Hamiltonian-Schwinger}
  \hat H=-\hbar\Omega_\mathrm{A} \hat{J}_{\mathrm{A}x}+\hbar U_\mathrm{A} \hat{J}_{\mathrm{A}z}^2-\hbar\Omega_\mathrm{B} \hat{J}_{\mathrm{B}x}+\hbar U_\mathrm{B} \hat{J}_{\mathrm{B}z}^2+2\hbar U_\mathrm{AB}\hat{J}_{\mathrm{A}z}\hat{J}_{\mathrm{B}z}.
 \end{align}

Let us first consider the interaction-free case $U_\mathrm{A}=U_\mathrm{B}=U_{AB}=0$. Under this condition,  both spins precess independently around the $x$-axis with angular velocities $\Omega_\mathrm{A}$ and $\Omega_\mathrm{B}$, respectively. 
The propagator can be expressed in terms of the matrix elements of SU(2) rotation operators, which are given (up to a phase factor) by the Wigner $d$-functions \cite{wigner_group_2013},
\begin{align}
\braket{k,l|\exp(-i\hat Ht/\hbar)|k_0,l_0}&= \braket{j_\mathrm{A},m_\mathrm{A}| \exp(i\Omega_\mathrm{A} \hat J_{\mathrm{A}x} t ) |j_\mathrm{A},m_{\mathrm{A}0}}\braket{j_\mathrm{B},m_\mathrm{B}| \exp(i\Omega_\mathrm{B} \hat J_{\mathrm{B}x} t ) |j_\mathrm{B},m_{\mathrm{B}0}}\\ &=  d^{j_\mathrm{A}}_{m_\mathrm{A},m_{\mathrm{A}0}}(\Omega_\mathrm{A} t)\  d^{j_\mathrm{B}}_{m_\mathrm{B},m_{\mathrm{B}0}}(\Omega_\mathrm{B} t)\exp\left[ i (m_{\mathrm{A}0}+m_{\mathrm{B}0}-m_\mathrm{A}-m_\mathrm{B}) \pi/2\right]\, .
\end{align}
Given the initial state $\ket{k_0,l_0}$, it follows that the probability to detect the state $\ket{k,l}$ after an evolution time $t$ factorizes into two independent probabilities associated with each particle type:
\begin{align}\label{prob-nonint}
p_{k,l}(t)=|d^{j_\mathrm{A}}_{m_\mathrm{A},m_{\mathrm{A}0}}(\Omega_\mathrm{A} t)|^2  |d^{j_\mathrm{B}}_{m_\mathrm{B},m_{\mathrm{B}0}}(\Omega_\mathrm{B} t) |^2.
\end{align}
The factorization of $p_{k,l}(t)$ in \eqref{prob-nonint} reflects the fact that, in the absence of interactions, distinguishable species evolve independently from each other.
The probability distribution of the total imbalance, as defined in \eqref{totimbalance}, is then given by the convolution of the individual species' distributions:
\begin{align}\label{convo}
p_m(t)=\sum_{m_\mathrm{A}+m_\mathrm{B}=m} |d^{j_\mathrm{A}}_{m_\mathrm{A},m_{\mathrm{A}0}}(\Omega_\mathrm{A} t)|^2  |d^{j_\mathrm{B}}_{m_\mathrm{B},m_{\mathrm{B}0}}(\Omega_\mathrm{B} t) |^2.
\end{align}

The probability distribution for the imbalance of one species (say B) is periodic in time, with the period $2\pi/\Omega_\mathrm{B}$ of the spin precession. If all particles of that species are initialized in the left well (this maximum imbalance configuration corresponds to the spin up state, $m_\mathrm{B}=j_\mathrm{B}$), they will all be found in the right well (spin down state, $m_\mathrm{B}=-j_\mathrm{B}$) after a time $\pi/\Omega_\mathrm{B}$.
Accordingly, in the case of a single non-interacting particle 
type, the waveguide lattice of size $1\times (N+1)$ introduced in the previous section is known as the $J_x$ lattice and allows perfect state transfer: a field amplitude injected at one end of the lattice is perfectly transferred to the other end at time $\pi/\Omega_\mathrm{B}$ and returns to its initial position after a full period $2\pi/\Omega_\mathrm{B}$ \cite{christandl_perfect_2004,perez-leija_coherent_2013,perez-leija_perfect_2013}.
This illustrates the mutual equivalence of the double well and photonic waveguide representations of \eqref{Hamiltonian}. The time $\pi/(2\Omega_\mathrm{B})$ is of particular significance, since after that time,  the B-particles have evolved into a balanced superposition of the left and right modes. 
At this instant in time, the tunnelling barrier plays the role of a balanced beam splitter. As in the HOM experiment, this balanced distribution of amplitudes induces maximal contrast of the many-particle interference signal. The simple periodic behaviour and the direct connection to a beam splitter setup described above are lost when interactions are turned on. In the Schwinger picture, the intra-species interactions 
generate one-axis spin-squeezing terms $\propto \hat J_{\mathrm{Az(Bz)}}^2$ \cite{kitagawa_squeezed_1993}, while inter-species interactions $\propto \hat J_\mathrm{A z}\hat J_\mathrm{B z}$ induce a coupling of the two spins through their $z$-components.

The Schwinger representation is particularly useful under isospecific conditions, since the  Hamiltonian \eqref{Hamiltonian-Schwinger} of the interacting system can then be brought into the Lipkin-Meshkov-Glick form \cite{lipkin_validity_1965,tichy_dynamical_2012} for the total spin $\hat{\vec{J}}=\hat{\vec{J}}_\mathrm{A}+\hat{\vec{J}}_\mathrm{B}$:
 \begin{align}\label{Hiso}
  \hat H=-\hbar\Omega \hat{J}_x+\hbar U \hat{J}_z^2.
 \end{align}
This prompts us to introduce  the common eigenstates $\ket{j,m}$ of $\hat{J}^2$ and $\hat{J}_z$, which, with the help of the Clebsch-Gordan coefficients $C^{j_\mathrm{A},j_\mathrm{B},j}_{m_\mathrm{A},m_\mathrm{B},m}$, can be expressed in terms of the Fock states $\ket{j_\mathrm{A},m_\mathrm{A},j_\mathrm{B},m_\mathrm{B}}$:
  \begin{subequations}\label{coupled}
  \begin{gather}
 \hat{J}^2 \ket{j,m}= j(j+1) \ket{j,m}, \qquad \qquad 	\hat{J}_{z} \ket{j,m}= m\ket{j,m},\\
\ket{j,m}= \sum_{m_\mathrm{A}+m_\mathrm{B}=m}   C^{j_\mathrm{A},j_\mathrm{B},j}_{m_\mathrm{A},m_\mathrm{B},m} \ket{j_\mathrm{A},m_\mathrm{A},j_\mathrm{B},m_\mathrm{B}},
 \end{gather}
 \end{subequations}
 with the usual conditions $|j_\mathrm{A}-j_\mathrm{B}| \leq j\leq j_\mathrm{A}+j_\mathrm{B}$ and $-j \leq m\leq j$ \cite{cohen-tannoudji_quantum_1991}. 
 Since the isospecific Hamiltonian \eqref{Hiso} commutes with $\hat J^2$, $j$ is a good quantum number. Moreover, as implied by our notation, the eigenvalues $m$ of $\hat J_z$ correspond to the total population imbalance, and the moments of $\hat J_z$ coincide with those of the imbalance distribution: $\braket{\hat J_z^n(t)}=\sum_m m^n p_m(t)$.
The isospecific Hamiltonian \eqref{Hiso} has the same form as that of a single-component BJJ, as can be seen by setting $\Omega_\mathrm{B}=\Omega$, $U_\mathrm{B}=U$  and $\Omega_\mathrm{A}=U_\mathrm{A}=U_\mathrm{AB}=0$ in  \eqref{Hamiltonian-Schwinger}. Therefore, the states $\ket{j,m}$ follow the same evolution as states of $N=2j$  bosons of the same species, e.g. $\ket{j_\mathrm{A}=0,m_\mathrm{A}=0,j_\mathrm{B}=j,m_\mathrm{B}=m}$.
 In particular, states with maximal spin $j=j_\mathrm{A}+j_\mathrm{B}$ are symmetric under the exchange of all particles, so that they behave like $N=N_\mathrm{A}+N_\mathrm{B}$ indistinguishable bosons. 
In contrast, if $N_\mathrm{A}=N_\mathrm{B}$, there exists a singlet state $\ket{j=m=0}$ which exhibits trivial time evolution, because it belongs to the one-dimensional invariant subspace $j=0$. We will discuss these cases in more detail in Sec.~\ref{Sec:Cases}.

\section{Typical dimensions and parameters}
\label{Sec:Typical}

Let us now briefly address the experimental aspects of the above scenario, and consider some realistic parameters for a typical waveguide array.
For definiteness, we assume equal particle numbers $N_\mathrm{A}=N_\mathrm{B}=6$ and an isospecific Hamiltonian.
This means that our all-optical implementation of the two-species BJJ is a $7\times7$ waveguide array with isotropic evanescent 
couplings $\kappa^\mathrm{A}=\kappa^\mathrm{B}=\kappa$. 
We fix the longitudinal length of the lattice to $L=\unit[15]{cm}$. 

\begin{figure}[t]
\begin{tabular}{c c c c}
	 {\large \bf (a)} & {\large \bf (b)} &  {\large \bf (c)}& \\
\includegraphics{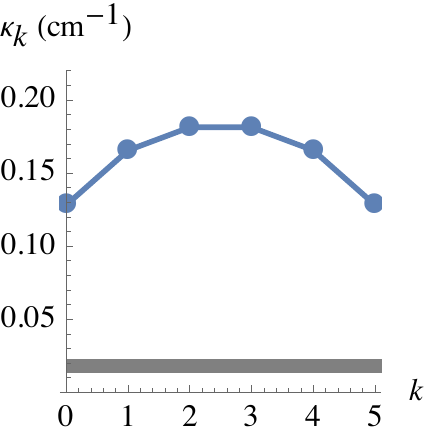} &\includegraphics{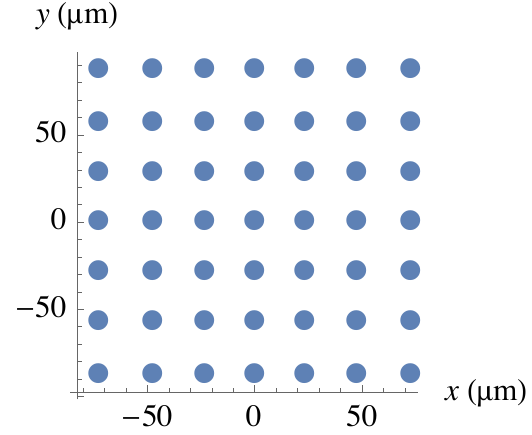}& \includegraphics{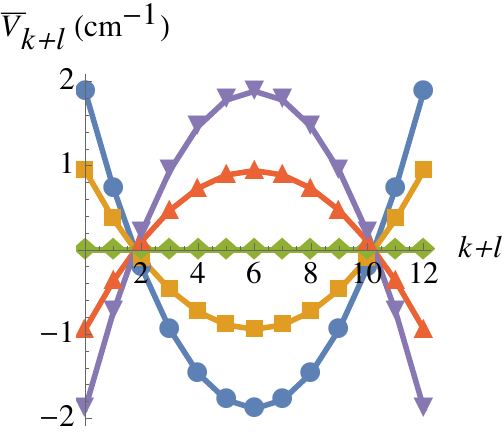}&\parbox[b][125pt][c]{50pt}{\includegraphics{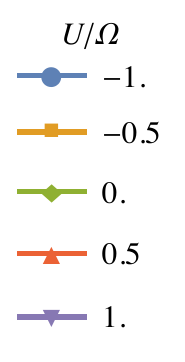}}
\end{tabular}
\caption{\label{Fig:Waveguides} 
Exemplary implementation of a $7\times7$ waveguide lattice to simulate the isospecific BJJ Hamiltonian with $N_\mathrm{A}=N_\mathrm{B}=6$ and $\Omega=2\pi/L=\unit[0.105]{cm^{-1}}$: \textbf{(a)} Distribution of inter-waveguide coupling constants $\kappa_k$, 
according to Eq.~\eqref{coupling}. The strength of the undesired diagonal couplings is indicated by the horizontal grey bar, and well-separated from horizontal/vertical coupling strengths. \textbf{(b)} Transverse waveguide arrangement with physical distances, given the vertical/horizontal coupling constant distribution of (a).
\textbf{(c)} On-site energies as given by Eq.~\eqref{detunings-shifted}, for various values of the interaction strength $U_\mathrm{A}=U_\mathrm{B}=U_\mathrm{AB}=U$. 
}
\end{figure}

As explained in the previous section, maximal interference contrast can be expected 
after an evolution time $T=\pi/(2\Omega)$, when the junction acts as a balanced beam splitter. Because evolution time is tantamount to propagation length along the waveguides, as established by the
mapping introduced in Sec.~\ref{Sec:Model}, we adjust $\Omega_\mathrm{A}=\Omega_\mathrm{B}=\Omega=\pi/(2L)\approx\unit[0.105]{cm^{-1}}$ in expression \eqref{coupling} for the couplings $\kappa$.  
We furthermore need to take into account that the evanescent coupling between two waveguides decays exponentially with their distance $d$, and that this decay may be anisotropic \cite{szameit_light_2007}.
To be specific, we set
\begin{equation}
\kappa^\mathrm{A}(d)=C_\mathrm{A} \exp(-\alpha_\mathrm{A} d),\qquad \kappa^\mathrm{B}(d)=C_\mathrm{B} \exp(-\alpha_\mathrm{B} d),
\label{distancescaling}
\end{equation}
with $C_\mathrm{A}=\unit[20]{cm^{-1}}$, $C_\mathrm{B}=\unit[30]{cm^{-1}}$, $\alpha_\mathrm{A}=\unit[0.20]{\textrm{\textmu m}^{-1}}$
and $\alpha_\mathrm{B}=\unit[0.18]{\textrm{\textmu m}^{-1}}$, which are typical parameters for laser-inscribed waveguides in fused silica glass
for illumination with $\lambda=\unit[633]{nm}$ light \cite{szameit_light_2007,Keil:GlauberFockLattices}.
The numerical values of the couplings of the $k$-th to $k+1$-th waveguide (recall \eqref{TightBinding}) are given by Eq.~\eqref{coupling} and are plotted in Fig.~\ref{Fig:Waveguides} \textbf{(a)}, together with the 
corresponding waveguide distances in Fig.~\ref{Fig:Waveguides} \textbf{(b)}.

Next, we need to determine the local detunings $V_{k,l}$, which -- by virtue of \eqref{detuning} -- only depend on the total number of particles on each site in the
isospecific case considered here:
\begin{equation}
V_{k,l}=V_{k+l}= -\frac{U}{2}\left[(k+l)^2+\left(N-k-l\right)^2-N \right]\, .
\end{equation}
We now take advantage of the fact that an arbitrary global 
shift of the detuning function does not affect 
the propagation dynamics, except for adding a global phase. Since minimal detuning from the reference value $0$ over the entire lattice 
is favourable for the calibration of the waveguide fabrication parameters (because then moderate variations of the inscription velocity can 
be used to set the detuning without affecting the coupling too strongly \cite{szameit_discrete_2010}), we can implement 
shifted detuning profiles as shown in 
Fig.~\ref{Fig:Waveguides} \textbf{(c)}, with  
\begin{equation}\label{detunings-shifted}
\bar{V}_{k+l}\equiv V_{k+l} - \frac{\max(V)+\min(V)}{2}=-\frac{U}{2}\left[(k+l)^2+\left(N-k-l\right)^2-\frac{3}{4}N^2\right].
\end{equation}

Finally, note that in a physical waveguide lattice one does not only encounter horizontal and vertical evanescent coupling, as intended by our model, but also some undesired next-to-nearest-neighbour coupling along the diagonals between sites $(k,l)$ and $(k\pm 1,l\pm 1)$. We approximate the strength of these couplings by an exponential dependence as in Eq.~\eqref{distancescaling}, with a characteristic range comparable to those of the horizontal and vertical coupling. 
The ratio of diagonal to horizontal (or vertical) coupling scales as $\kappa(\sqrt{2}d)/\kappa(d)\sim e^{-\alpha(\sqrt{2}-1) d}$, 
and is therefore minimized if well-separated waveguides are used (this in turn implies long waveguides if $\Omega L$ is kept constant).
With the above choice of parameters, the ratio between diagonal and horizontal (or vertical) couplings is on the order of 10\% (see Fig.~\ref{Fig:Waveguides} \textbf{(a)}). To assess deviations between experimental results and the predictions of the model \eqref{TightBinding}, in the following we 
provide simulations which both include and neglect diagonal couplings.

\section{Cases of interest}
\label{Sec:Cases}

The above optical implementation of the two-component BJJ allows us to systematically assess the role of (in)distinguishability in many-body dynamics, by direct comparison of the dynamics when only the distinguishability of the participating particles is changed. Specifically, we now compare the probability distributions $p_m(t)$ for various initial states with the same initial total imbalance distribution $p_m(0)$ but different distributions of the particle species. We again focus on the isospecific case, such that differences in the resulting dynamics can be unambiguously attributed to many-body interference effects.

\begin{figure}[t]
	
	\begin{tabular}{c  c p{5pt} c p{5pt} c}
 \parbox[b][140pt][c]{50pt}{ {~\vspace{55pt} \\ \large \bf (a)} \\~\\ \includegraphics[width=50pt]{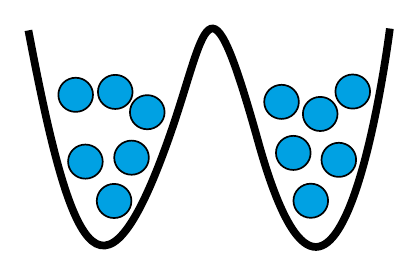} \\
 \includegraphics[width=50pt]{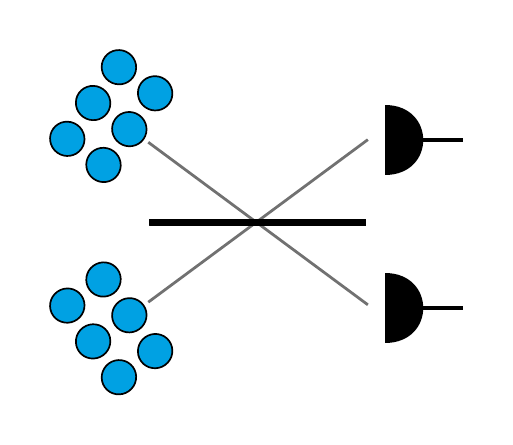}}	& \parbox[b][140pt][c]{140pt}{\includegraphics{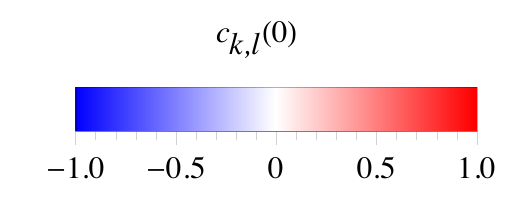}\vfill 
		\includegraphics{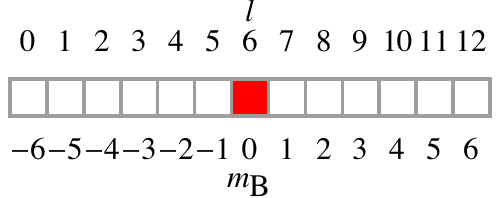}}
	& &\parbox[b][140pt][c]{140pt}{\includegraphics{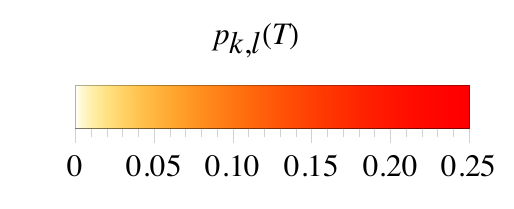} \vfill 
		\includegraphics{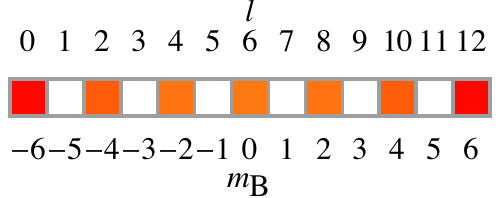} }
	&  &\includegraphics{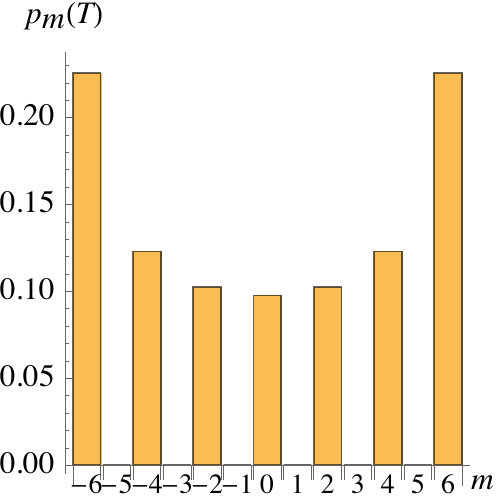}\\[5pt] 
 \parbox[b][140pt][c]{50pt}{ {\large \bf (b)} \\~\\ \includegraphics[width=50pt]{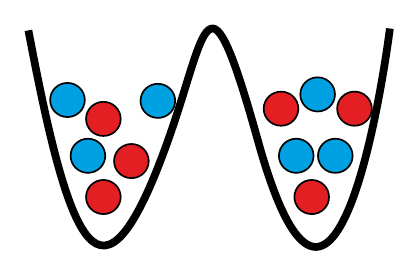} \\
 	\includegraphics[width=50pt]{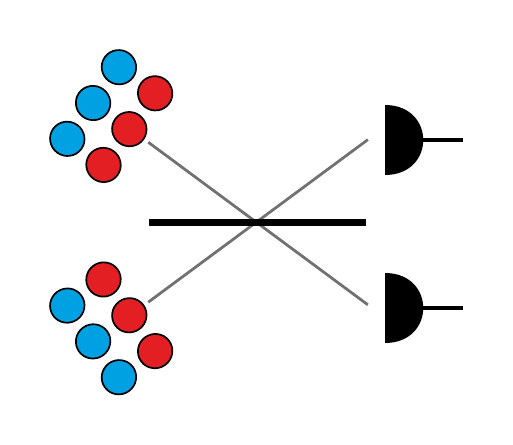}}	&\includegraphics{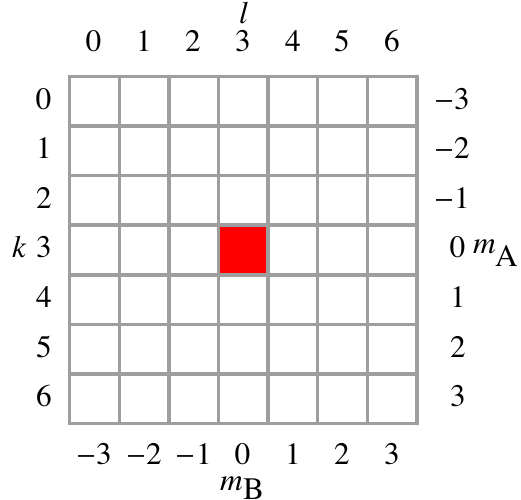} && \includegraphics{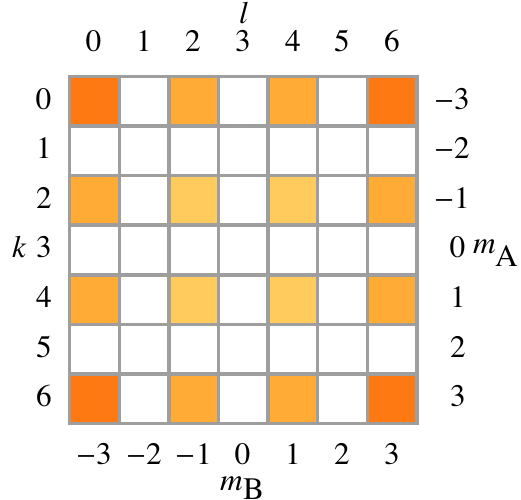} && \includegraphics{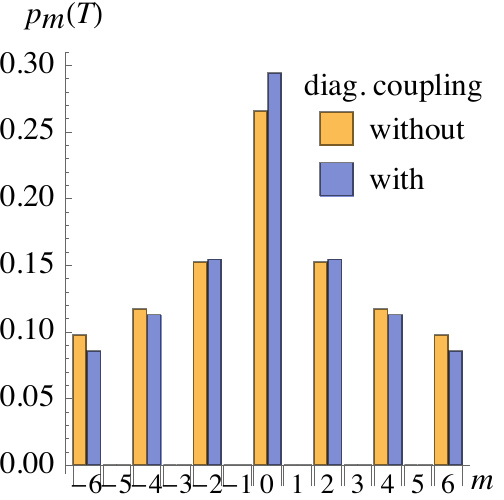}	\\[5pt]   
 \parbox[b][140pt][c]{50pt}{ {\large \bf (c)} \\~\\ \includegraphics[width=50pt]{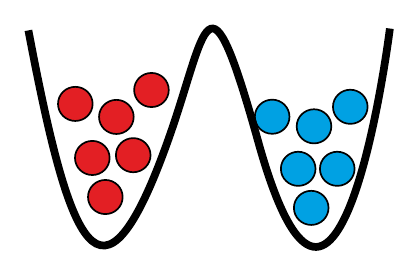} \\
 	\includegraphics[width=50pt]{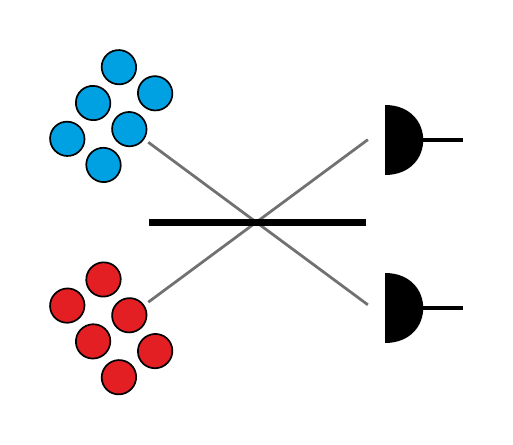}}	&\includegraphics{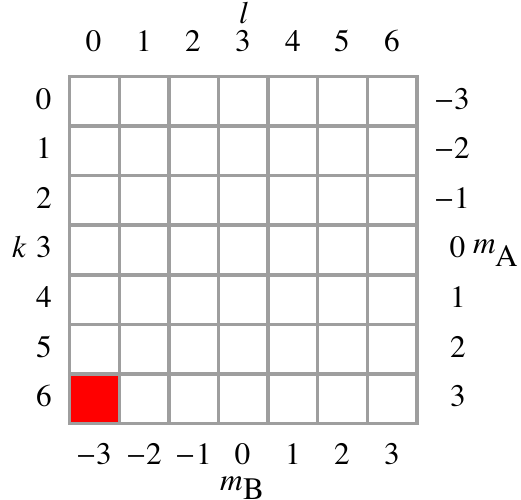} && \includegraphics{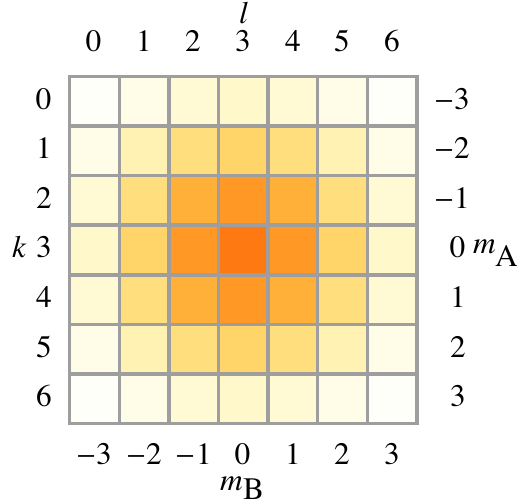} && \includegraphics{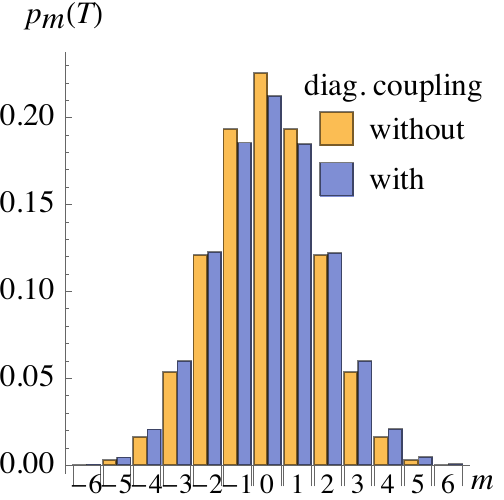}
\end{tabular}	
	
	\caption{\label{Fig:NonInt} Initial amplitude distribution, $c_{k,l}(0)$ (first column), and final intensity distribution, $p_{k,l}(T=\pi/(2\Omega))$ (central column), of the field over the waveguide array, as generated by \eqref{TightBinding} for vanishing interaction strengths $U_\mathrm{A}=U_\mathrm{B}=U_\mathrm{AB}=0$ in the BJJ model \eqref{Hamiltonian}. The right column displays the imbalance distribution $p_m(T)$ as defined in \eqref{totimbalance}, both for the idealized model (neglecting undesired diagonal couplings between the waveguides) and taking non-vanishing diagonal couplings into account, for parameter values as described in Sec.~\ref{Sec:Typical}. It is apparent that typical diagonal couplings do not affect the essential physics on the time scales considered here. Top, middle and bottom row represent \textbf{(a)} the single-species case, \textbf{(b)} the mixed initial state with $N_A=N_B=N/2$ particles of each species equally distributed over both wells, and \textbf{(c)} the separated initial state with each particle species fully localized in one well, respectively.}
\end{figure}

In the following, we consider initial states with imbalance $m_0=k_0+l_0-N/2=0$, that is $p_m(0)=\delta_{m,0}$, where $\delta$ is the Kronecker delta.
Our reference dynamics is defined by the case where {\it all} particles belong to the {\it same} species (say B), as illustrated in Fig.~\ref{Fig:NonInt} \textbf{(a)}. 
In our photonic implementation, this corresponds to an excitation at the centre of a one-dimensional waveguide array with $N+1=N_B+1$ sites (left panel in row \textbf{(a)} of Fig.~\ref{Fig:NonInt}). 
To compare this to two-species scenarios, we consider the case $N_\mathrm{A}=N_\mathrm{B}=N/2$, which corresponds to a square array of waveguides (like the situation depicted in Fig.~\ref{Fig:Waveguides}).
In one extreme case, both particle types are equally distributed over the wells, that is $N_\mathrm{A}/2$ A-particles start in the left mode and $N_\mathrm{A}/2$ in the right mode, and the same holds for B, as represented in Fig.~\ref{Fig:NonInt} \textbf{(b)}. This corresponds to the initial state $c_{N/4,N/4}(0)=1$, i.e., a single waveguide excitation in the centre of the structure (left panel in row \textbf{(b)} of Fig.~\ref{Fig:NonInt}). 
In the opposite limit, particles from different species are initially completely separated, with all A-particles in the left mode and all B-particles in the right, as represented in Fig.~\ref{Fig:NonInt} \textbf{(c)}. This corresponds to an excitation of a corner of the lattice: $c_{N_\mathrm{A},0}(0)=1$ (left panel in row \textbf{(c)} of Fig.~\ref{Fig:NonInt}). Hereafter we will call these initial states \textit{mixed} (b) and \textit{separated} (c), respectively.

\begin{figure}[t]

	\begin{tabular}{c  c p{5pt} c p{5pt} l}

		& \includegraphics{legendbluewhitered}&&	 \includegraphics{legendred4} && \\ 
	 \parbox[b][140pt][c]{20pt}{ \large \bf (a) }	&\includegraphics{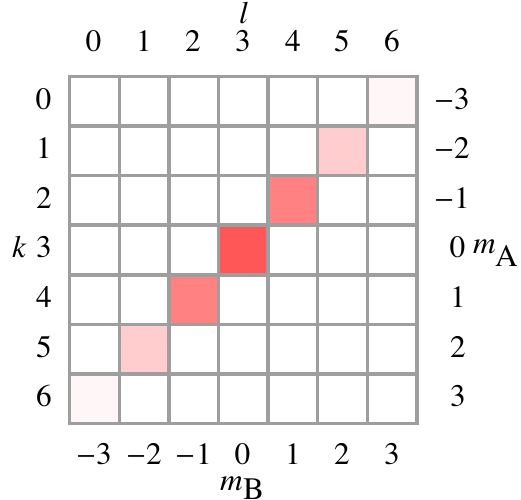} && \includegraphics{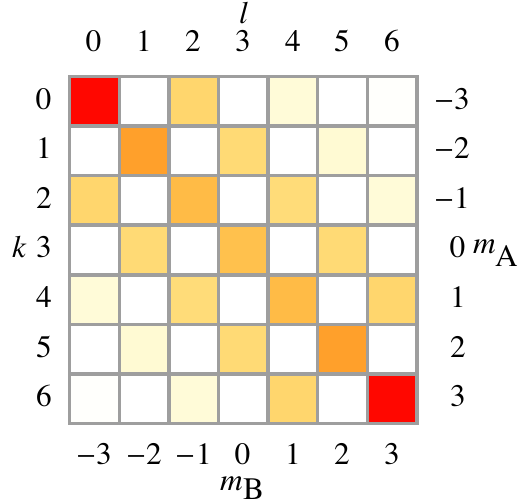} && \includegraphics{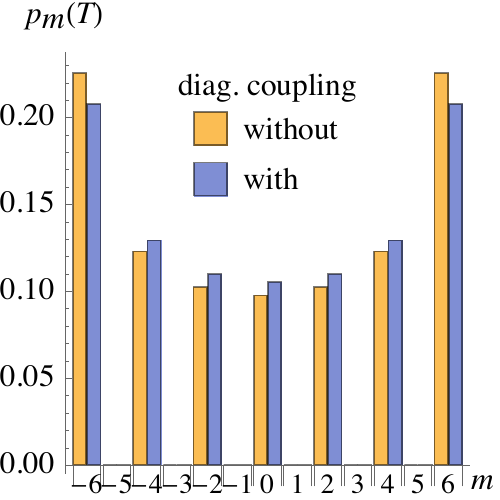}	\\[5pt] 
	 \parbox[b][140pt][c]{20pt}{ \large \bf (b) }	&\includegraphics{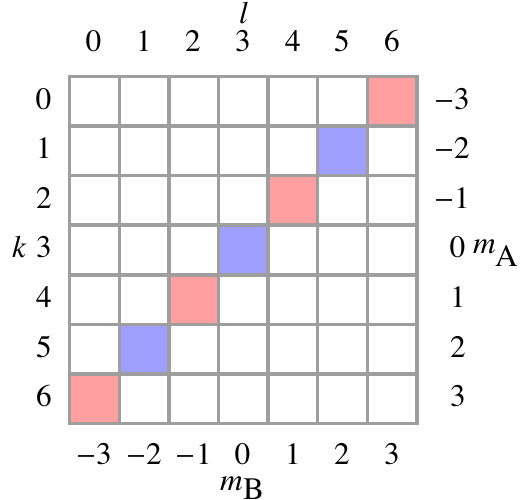} && \includegraphics{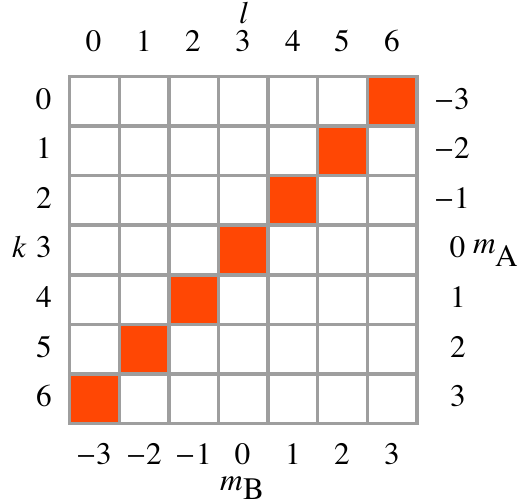} && \includegraphics{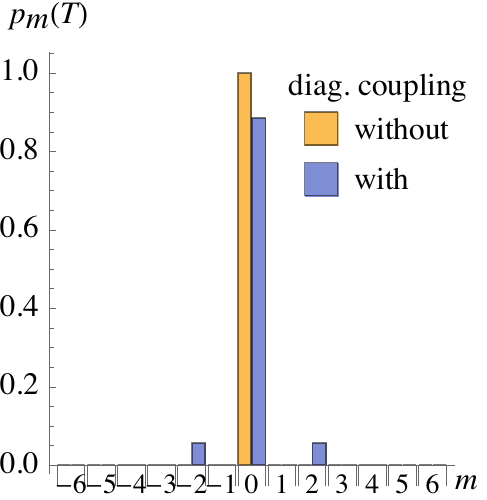}
	\end{tabular}	
	
	\caption{\label{Fig:NonIntCoupled} Same as Fig.~\ref{Fig:NonInt}, now for initial amplitude distributions $c_{k,l}(0)$ representing eigenstates $\ket{j,m}$ of the total spin \eqref{coupled} in the Schwinger representation \eqref{Hiso} of the isospecific BJJ model. While \textbf{(a)} $\ket{j=N/2, m=0}$ precisely mimics the dynamics of the single-species scenario depicted in Fig.~\ref{Fig:NonInt} (a), \textbf{(b)} $\ket{j=0, m=0}$ does not evolve, due to its vanishing total spin.
		}
\end{figure}

Finally, one can reproduce single-species dynamics in the 2D waveguide lattice by initializing the system in an eigenstate $\ket{j,m}$ of the total spin, as explained in Sec.~\ref{Sec:Schwinger}. This requires to simultaneously inject light in those waveguides which are located on the antidiagonal with $m_\mathrm{A}+m_\mathrm{B}=m$, with amplitudes given by the Clebsch-Gordan coefficients introduced in Eqs.~\eqref{coupled}.
We here consider two exemplary cases: First, the balanced, fully symmetric state, with $j$ equal to its maximum value $j=N/2$ and $m=0$, which corresponds to initial amplitudes $c_{k,l}(0)=\delta_{k+l,N/2} \sqrt{\binom{N}{N/2}}\binom{N/2}{k}$ (left panel in row \textbf{(a)} of Fig.~\ref{Fig:NonIntCoupled}), and, second, the state with $j=m=0$, with amplitudes  $c_{k,l}(0)=\delta_{k+l,N/2} (-1)^{k}/\sqrt{N/2+1}$ (left panel in row \textbf{(b)} of Fig.~\ref{Fig:NonIntCoupled}). 

We now inspect the resulting dynamics for $N=12$ particles, which corresponds to a $1\times 13$ lattice in the single-species case, and to a  $7\times 7$ lattice in the two-species case, with the parameters as determined in Sec. \ref{Sec:Typical}. In all cases, the final imbalance distributions were calculated with and without residual diagonal couplings between the waveguides. As we will see, there is some influence of the diagonal coupling for the chosen parameter values, which, however, does not affect the general physical picture and can therefore be tolerated.

\subsection{Interaction-free case}

Let us start with the non-interacting case $U_\mathrm{A}=U_\mathrm{B}=U_\mathrm{AB}=0$.
The various initial configurations are given in the left column of Figs.~\ref{Fig:NonInt} and \ref{Fig:NonIntCoupled}, while the corresponding probability distributions at time $T=\pi/(2\Omega)$ are shown in the middle and right columns.
In this case, the initial and final states match the input and output of a balanced beam splitter,
it is therefore insightful to interpret these results in terms of many-particle interference: In the single-species scenario (row \textbf{(a)} in Fig.~\ref{Fig:NonInt}), all particles are indistinguishable from each other. 
Therefore, the central site excitation, corresponding to $N/2$ particles in either mode initially, is equivalent to sending $N$ bosons symmetrically onto a balanced beam splitter \cite{campos_quantum-mechanical_1989,stobinska_quantum_2015}. It is well known that strong many-particle interference arises in this setup, which allows only even numbers of particles in either mode and renders bunched configurations with all particles ending on one site the most likely outcomes \cite{hong_measurement_1987,Ou:FourPhotonsBS,Niu:SixPhotonBSBunching,Ra:DetectiondependentCoherenceTimes}. This is clearly reflected in the imbalance distribution (right panel), which exhibits non-vanishing probabilities only for even $m$ and the strongest signal at $m=\pm N/2$. 

For the mixed input state (row  \textbf{(b)} in Fig.~\ref{Fig:NonInt}), one has now $N/4$ particles of each type in each mode. Particles of the same species still undergo the aforementioned bosonic interference, enforcing an output configuration with even numbers of A- and B-particles in both modes, while odd numbers remain forbidden. 
However, neither interaction nor interference occur between particles of different types. Therefore, the combined imbalance distribution (right panel) is the convolution \eqref{convo} of the individual distributions, which leads to the central peak.

For the separated input state (row \textbf{(c)} in Fig.~\ref{Fig:NonInt}), no many-particle interference occurs whichsoever, because all A-particles (respectively B-particles) start on the same site.
Therefore, one ends up with a simple binomial distribution which reflects that all particles evolve independently (right panel).

Finally, we consider the eigenstates $\ket{j=N/2, m=0}$ and $\ket{j=0, m=0}$ of the coupled spins.
As expected, the former (row  \textbf{(a)} in Fig.~\ref{Fig:NonIntCoupled}) behaves exactly like the corresponding single-species state (compare with case \textbf{(a)} in Fig.~\ref{Fig:NonInt}). In contrast, the latter (row  \textbf{(b)} in Fig.~\ref{Fig:NonIntCoupled}) does not evolve at all, since it corresponds to a state with zero total spin, which does not precess.

\subsection{Dynamics with interaction}

\begin{figure}[t]

\setlength{\tabcolsep}{10pt}
	\begin{tabular}{ c  c  c}
		{ \large \bf (a) } & { \large \bf (b) } & { \large \bf (c) }\\
	 \includegraphics{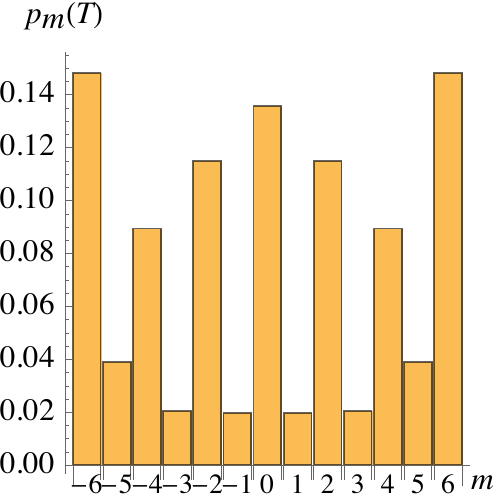}&	 \includegraphics{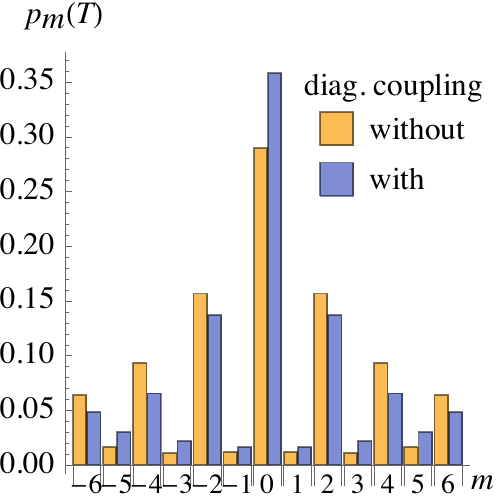} & \includegraphics{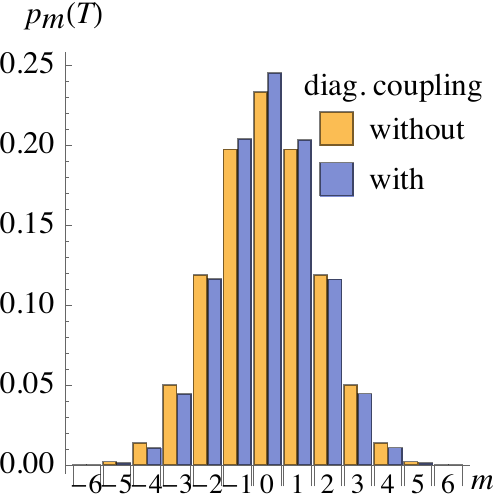} \\[5pt] 
	 \includegraphics{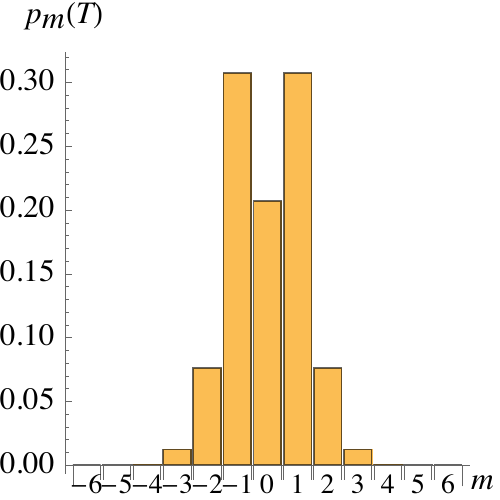} & \includegraphics{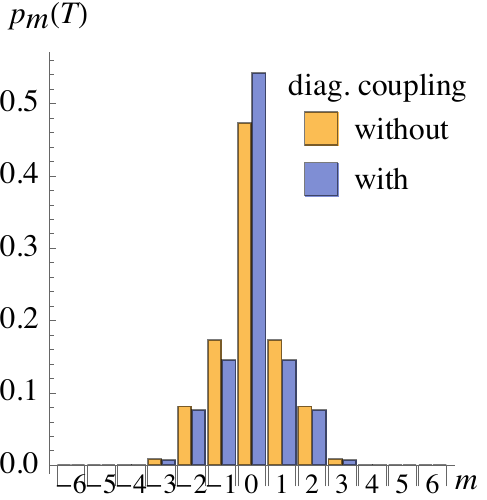}& \includegraphics{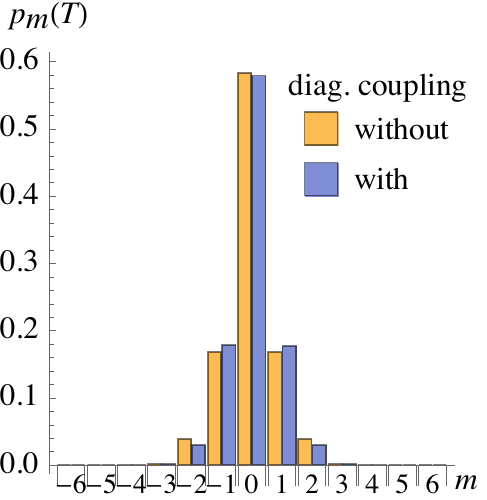}
\end{tabular}	

\caption{\label{Fig:Int} 
Imbalance distributions $p_m(T)$ at final time $T=\pi/(2\Omega)$, for moderate (top, $U=0.125\ \Omega$)
and strong (bottom, $U=\Omega$) interactions $U_A=U_B=U_{AB}=U$, with a comparison of the results for vanishing and non-vanishing diagonal couplings in the two-species scenarios. Initial 
states as cases (a,b,c) in Fig.~\ref{Fig:NonInt}. The resulting distributions' contrast is decreasing with increasing $U$, and the distributions tend to concentrate
around $m=0$, for strong interactions.}
\end{figure}

We now turn on the interactions $U_\mathrm{A}=U_\mathrm{B}=U_\mathrm{AB}=U\neq 0$. In this case, the equivalence with photons interfering on a beam splitter no longer holds, but the mapping to the coupled waveguide system is still valid. Moderate interactions induce a dephasing of the tunnelling oscillations, 
with the result that the interference features observed in the interaction-free case are progressively washed out. However, the time at which the interference contrast is maximal remains the same as in the non-interacting case, i.e., $T=\pi/(2\Omega)$.
The degradation of the many-particle interference visibility can be observed in the top row of Fig.~\ref{Fig:Int}, for $U=0.125\ \Omega$: in the single-species case as well as for two species prepared 
in the mixed input state, the even-odd interference pattern has lost contrast.
Moreover, all distributions become more peaked towards the centre. This can be understood from the fact that interactions  tend to 
suppress tunnelling by increasing the energy difference between states with different population imbalances, and 
the effect is more and more pronounced with increasing interaction strengths, as witnessed in the bottom row of Fig.~\ref{Fig:Int}, for $U=\Omega$. Consistently, with the progressive suppression of many-particle interference effects the initial state preparation has decreasing impact on the final  imbalance distribution.

\begin{figure}[t]
\includegraphics{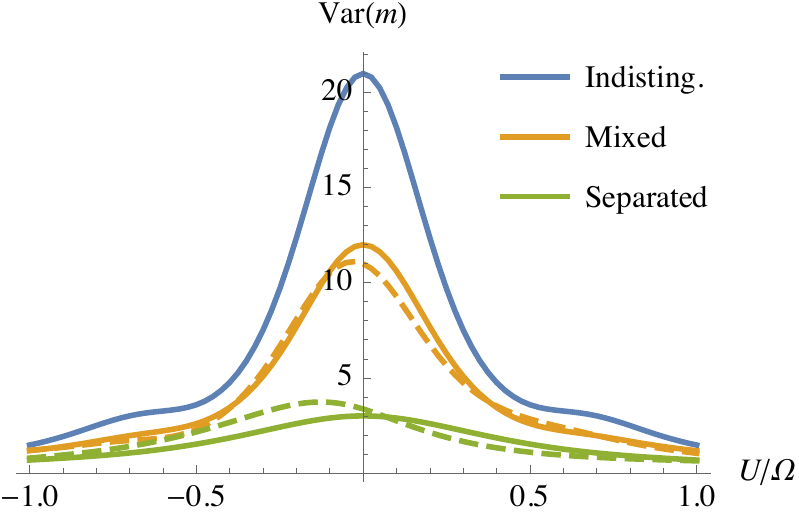}
\caption{\label{Fig:Variances} Variance $\text{Var}(m)$ of the imbalance distribution $p_m(t)$, at the final time $t=T=\pi/2\Omega$, as 
a function of the (attractive or repulsive) interaction strength $U$. The full blue line presents the result for the single-species initial state. Full (dashed) lines indicate the expected results in the absence (presence) of diagonal couplings, for separated and mixed two-species initial states (yellow and green lines, respectively).  }
\end{figure}

Given the symmetry of the problem, the average imbalance $\bar{m}=\sum_m m p_m(t)$ is zero in all the above cases (with and without interaction), except for non-vanishing diagonal couplings, which can break the symmetry in conjunction with the $x$-$y$-coupling anisotropy, and therefore give rise to slight deviations.
The various scenarios can, however, easily be distinguished by the spread of the distribution, as measured by the variance $\mathrm{Var}(m)=\sum_m m^2 p_m(t)$. 
The variance of the distribution at time $T$ reflects the level of interference, as can be seen in Fig.~\ref{Fig:Variances}: the single-species distribution has the largest variance due to its pronounced outer peaks ($\mathrm{Var}(m)=21$ for $U=0$), the state with initially separated species leads to a much smaller value ($\mathrm{Var}(m)=3$ for $U=0$), and the variance of the distribution for a mixed initial state lies between those two cases ($\mathrm{Var}(m)=12$ for $U=0$).
 As the interaction strength increases, the variances decrease sharply for states featuring many-particle interference until they become comparable to that of the separated state, which does not allow any interference.

Note that the above considerations are valid both for attractive and repulsive interactions, since both cases are related by a symmetry of the Hubbard model under time-reversal \cite{schneider_fermionic_2012,ronzheimer_expansion_2013,lahini_quantum_2012}. We briefly outline the argument: 
Define the involutive antiunitary operator $\hat \Pi=\hat K \hat P$, where  $\hat K$ denotes complex conjugation in the Fock basis and corresponds to a time-reversal operation, while 
$\hat P$ is a redefinition of the phases of the Fock basis vectors through $\hat P\ket{k,l}=(-1)^{k+l}\ket{k,l}$. Under the action of $\hat P$, the hopping term in the Hamiltonian \eqref{Hiso} changes sign while the interaction term is invariant. Since the Hamiltonian is real in the Fock basis, $\hat \Pi$ has the same effect. Therefore, if $\hat H(U)$ denotes the Hamiltonian for an interaction strength $U$, then $\hat \Pi \hat H(U) \hat \Pi = - \hat H(-U)$ and the corresponding evolution operators are related by $\hat \Pi \exp(-i \hat H(U)t) \hat \Pi = \exp(-i \hat H(-U)t)$. It follows that for any observable $\hat O$ with $\hat \Pi \hat O\hat \Pi=\hat O$, the expectation value in a Fock state at time $t$ is independent of the sign of $U$. This is in particular the case for $\hat O=\hat J_z$ and $\hat J_z^2$, from which we conclude that the width of the imbalance distribution is invariant upon changing the sign of interactions, as is evident from the symmetry of the curves in Fig.~\ref{Fig:Variances}. A slight difference between the attractive and repulsive cases arises in the presence of diagonal coupling, which breaks the aforementioned symmetry.

\section{Conclusion}
\label{Conc}

By constraining the symmetry of the many-body wavefunction, indistinguishability deeply affects the dynamics of many-body systems. Splitting the particles between two distinguishable species relaxes this constraint and leads, even in the case of identical Hamiltonian parameters for both species, to strikingly different behaviours, which depend on the precise repartition of the different particle types.
We investigated this effect for interacting bosons in a double-well potential, and established that this scenario 
can be simulated with a lattice of coupled optical waveguides. Introducing two distinguishable species in the bosonic system amounts to increasing 
the dimension of the waveguide array from one to two, giving a simple geometrical interpretation to an otherwise puzzling effect. 
Another enlightening perspective is obtained using the Schwinger spin picture, where the dynamics of a single species is represented by a 
single spin, while in the two-species scenarios one must consider the sum of two spins.
Using the theory of addition of angular momenta, one can then build states where the particles are effectively indistinguishable or form 
an invariant eigenstate of the system, by a suitable superposition of two-species states.
In the non-interacting case, the final distribution of particles between the wells can be understood in terms of generalized HOM interference. 
In particular, for balanced inputs of indistinguishable particles, the state at time $T=\pi/(2\Omega)$ exhibits a complete suppression of output events with an odd number of particles in one of the wells, while those with all particles in the same well are enhanced.  
This interference is progressively destroyed as one renders the particles distinguishable, leading to a purely classical binomial distribution of particles in the extreme case of initially separated species. Weak attractive or repulsive interactions lead to dephasing which reduces the contrast of the interference patterns and affects their 
structure. Stronger interactions suppress tunnelling, resulting in a reduced variance  and a diminished impact of (in)distinguishability on the imbalance distribution.

Our numerical results show that the discussed phenomena can be experimentally investigated in 2D waveguide lattices with parameters typical for state-of-the-art laser-written waveguides. This allows to study the interplay of indistinguishability and interactions in a model system requiring only non-interacting photons in bright coherent states. Next-nearest-neighbour coupling, which is an intrinsic effect of the model system without a counter-part in the BJJ, does slightly influence the results, but is inconsequential for the general physical trends at realistic parameter values as chosen here. To improve the accuracy of the model system, these effects can be asymptotically eliminated by using larger waveguide separations.

The same platform can be used to study other aspects of the two-component BJJ model. Here, we have focused on  input states with the same number of particles in each well, which display the highest level of many-particle interference, but arbitrary Fock states or superpositions thereof can in principle be simulated.
Moreover, the transition between the case of all particles belonging to a single species  and that of particles equally split between the two species can be studied with the help of rectangular waveguide lattices, corresponding to $N_\mathrm{A}\neq N_\mathrm{B}$. By suitably choosing the lattice parameters, one may also implement non-isospecific dynamics, and in particular investigate the competition of inter- and intra-species interactions.
Another natural extension of our work is to consider a BJJ with more than two species. This can no longer be emulated in a waveguide lattice but it is conveniently mapped to a system of many interacting spins in the Schwinger picture. 

\section*{Acknowledgements}

	The authors thank Alexander Szameit for fruitful discussions.
	G.D. and A.B. express gratitude to the EU Collaborative project QuProCS (Grant Agreement No. 641277) for financial support. T.B. acknowledge support by the German Research Foundation (IRTG 2079). C.D., G.W. and R.K. are grateful to the Austrian Science Fund (project M 1849).

\end{document}